\documentclass{iaus}
\usepackage{graphicx}

\usepackage{xspace}

\usepackage{color}
\usepackage{ulem}

\newcommand{\phiob}{$\Phi_{\mathrm{OB}}$\xspace}

\newcommand{\phiholmes}{$\Phi_{\mathrm{HOLMES}}$\xspace}
\newcommand{\phitot}{$\Phi_{\mathrm{total}}$\xspace}

\newcommand{\Ha}{H$\alpha$\xspace}
\newcommand{\Hb}{H$\beta$\xspace}

\newcommand{\hii}{H~{\sc ii}}

\newcommand{\nii}{[N~{\sc ii}]}

\newcommand{\oii}{[O~{\sc ii}]}

\newcommand{\oiii}{[O~{\sc iii}]}

\newcommand{\sii}{[S~{\sc ii}]}

\title[Ionization of the diffuse gas in galaxies] 
{Ionization of the diffuse gas in galaxies : \\
 Hot low-mass evolved stars at work}

\author[Flores-Fajardo et al.]   
{N. Flores-Fajardo$^1$\thanks{E-mails:NahieFlores@Gmail.com, Chris.Morisset@Gmail.com, grazyna.stasinska@obspm.fr},
  C. Morisset$^{2,3}$, G. Stasi\'nska$^4$, L. Binette$^2$\\
}

\affiliation{$^1$ Centro de Radioastronom\'{\i}a y Astrof\'{\i}sica,
  Universidad Nacional Aut\'onoma de M\'exico; 
Antigua Carretera a P\'atzcuaro no. 8701 
Col. Ex Hacienda San Jos\'e de la Huerta, 
Morelia, Michoac\'an C.P. 58089, M\'exico\\[\affilskip]
$^2$Instituto de Astronom\'{\i}a,
     Universidad Nacional Aut\'onoma de M\'exico;
     Apdo. postal 70--264; Ciudad Universitaria;
     M\'exico D.F. 04510; M\'exico.\\ [\affilskip]
$^3$ Instituto de Astrof\'{\i}sica de Canarias;
C/ V\'{\i}a L\'actea, s/n,
E38205, La Laguna (Tenerife), Espa\~na\\[\affilskip]
$^4$LUTH, Observatoire de Paris, CNRS, Universit\'e Paris Diderot; Place Jules Janssen 92190 Meudon, France.}

\pubyear{2011} 
\volume{284}  
\pagerange{1--12}
\jname{The Spectral Energy Distribution of Galaxies}
\editors{R.J. Tuffs \&  C.C.Popescu, eds.}
\begin{document}

\maketitle

\begin{abstract}
We revisit the question of the ionization of the diffuse medium in late type galaxies, by studying NGC 891. The most important challenge for the models considered so far was the observed increase of \oiii/\Hb, \oii/\Hb, and \nii/\Ha\ with increasing distance to the galactic plane. We propose a scenario based on the \textit{expected }population of  massive OB stars and hot low-mass evolved stars (HOLMES) in this galaxy to explain this observational fact. 
In the framework of this scenario we construct a finely meshed grid of photoionization models. For each value of the galactic latitude $z$ we look for the models  which simultaneously fit the observed values of the \oiii/\Hb, \oii/\Hb, and \nii/\Ha  ratios.  For each value of $z$ we find a range of solutions  which depends on the value of the oxygen abundance. The models  which fit the observations indicate a systematic decrease of the electron density with increasing $z$. They become dominated by the HOLMES  with increasing $z$ only when restricting to solar oxygen abundance models, which argues that the metallicity above the galactic plane should be close to solar. They also indicate that N/O increases with increasing $z$.
\keywords {galaxies: individual (NGC 891) --- galaxies: ISM --- galaxies: abundances --- stars: AGB and post-AGB}
\end{abstract}

\firstsection 

\section{Introduction}

The Diffuse Ionized Medium (DIG) was detected through its optical line
emission outside the classical \hii\, regions
(\cite[Reynolds (1971)]{1971PhDT.........1R}) and turns out to be a
major component of the interstellar medium in galaxies (\cite[Reynolds
(1991)]{1991IAUS..144...67R}). Most
specialist agree that OB stars in galaxies likely represent the main
source of ionizing photons for the DIG (see \cite[Haffner et al. (2009)]{2009RvMP...81..969H}). However, an additional
ionizing source is suggested by reported increase of such emission
line ratios as \nii/\Ha, \sii/\Ha with galactic height. Several
sources of additional ionization/heating have been suggested without
complete success.

\section{Proposed Scenario}

\begin{figure}
\begin{center}
 \includegraphics[width=2.1in]{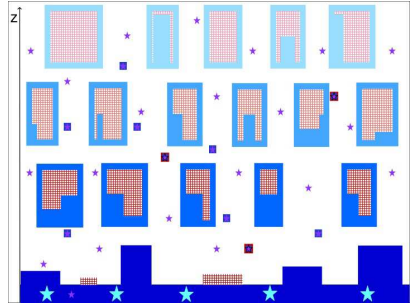} 
 \caption{Schematic representation of the extraplanar gas and ionizing
   stars. Ionized gas is represented in blue, neutral gas in red, with
   a lighter shade for less dense regions. For both components, a
   darker shade indicates gas of higher densty. The big blue stars in
   the plot represent OB stars. The small purple stars represent teh
   HOLMES, which can be either central satrs of present-day planetary
   nebulae, or hot pre-white dwarfs.}
   \label{fig1}
\end{center}
\end{figure}

The extraplanar DIG is destributed in clouds (for simplicity
represented by rectangles) that are ionized by two star populations:1)
OB stars located in the thin disk and whose ionizing radiation escape
from the disk through small ``holes'', which ionize the ``bottom''
part of the gas clouds. They constitute only a small fraction of the
entire population of the OB stars in the disk. 2) HOLMES, which are
distributed in the galaxy thick disk and halo. Their influence with
respect to that of OB stars increases away from the galacti
plane. Figure \ref{fig1} shows the datails of the proposed scenario.

\section{Observational Data}

\begin{figure}
\begin{center}
 \includegraphics[width=3.in]{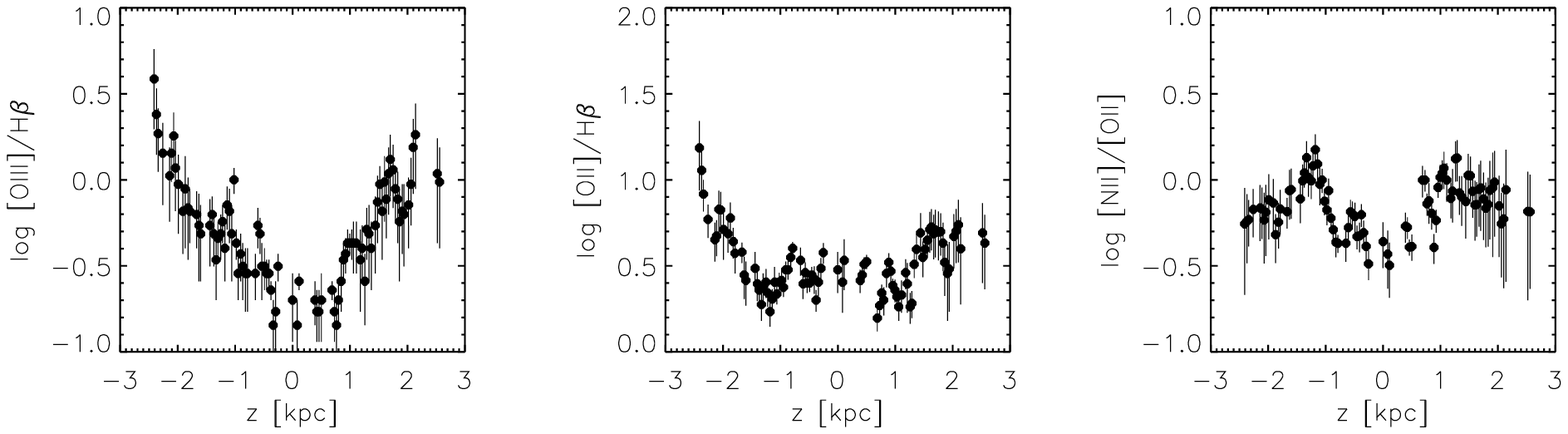} 
 \caption{Observed values of \oiii/\Hb, \oii/\Hb and \nii/\oii in NGC
   891, as a function of the distance to the galactic plane. The data
   are from \cite[Otte et al. (2001)]{2001ApJ...560..207O}.}
   \label{fig2}
\end{center}
\end{figure}

We chose to focus on the edge on spiral galaxy NGC 891, a
galaxy that has been extensively observed, especially in optical
emission lines (Figure \ref{fig2}), providing the best diagnostics for our
scenario.

\section{Modeling}

The ionizing spectral energy distribution (SED) from OB stars is
obtained using the code Starburst99 (\cite[Leitherer et al. (1999)]{1999ApJS..123....3L}), considering a continuous star
formation. The SED of HOLMES is obtained using the code PEGASE (\cite[Fioc 
\& Rocca-Volmerange (1997)]{1997A&A...326..950F})
considering a instantaneous starburst at look back after 10 Gyr. The
photoionization models for the DIG were computed with Cloudy (\cite[Ferland et al. (1998)]{1998PASP..110..761F}). Each
one is defined by the ratio of the surface fluxes: \phiholmes / (\phiob + \phiholmes), the ionization
parameter $U$, O/H and N/O.

\begin{figure}
\begin{center}
 \includegraphics[width=3.2in]{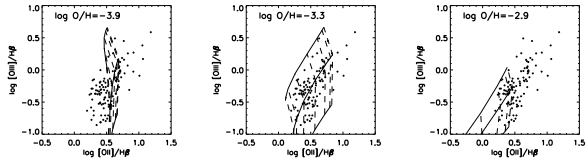} 
 \caption{``Reduced'' grids (the full grid is 5 time more resolved) of
   models in the \oiii/\Hb vs. \oii/\Hb plane, for various values of
   O/H. See text for details on the varying parameters. The
   observational points are superimposed on the grid.}
   \label{fig3}
\end{center}
\end{figure}

Figure \ref{fig3} shows the location of the observaional points in the
\oiii/\Hb vs \oii/\Hb diagram,
with respect to our ``reduced'' grid of models. The dashes lines join
models with same values of log \phiob (equal to 3.5, 4, 4.5, 5, 5.5, 6, 6.5),
while the continuous line join models with the same values of $U$ (equal
to -4, -3.5, -3). Form left to right, each panel corresponds to a
different value of log O/H (-3.9, -3.3, -2.9). The value of N/O is the same in
all the panels, namely -0.5 dex. The observational points that have
the highest values of \oiii/\Hb and \oii/\Hb correspond to the largest values of \phiholmes/\phitot.

\section{Results}

\begin{figure}
\begin{center}
 \includegraphics[width=4.6in]{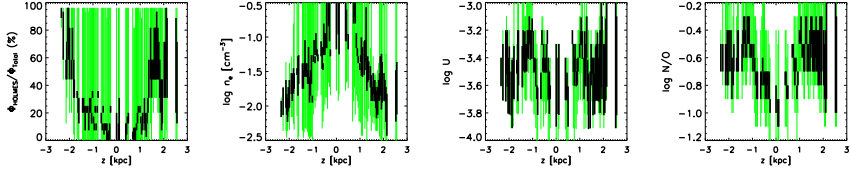} 
 \caption{\phiholmes/\phiob, $n_\mathrm{e}$, $U$ and N/O vs $z$ for
   the models that fit the observational \oiii/\Hb, \oii/\Hb, and
   \nii/\Ha simultaneously. The green bars indicate the range of the
   values for wich the models in the grid fit the observations. The
   dark bars show the same, but restricting to solar abundances models.}
   \label{fig4}
\end{center}
\end{figure}

For each observational point in the DIG of NGC 891 we select the
models of our finely meshed grid that reproduce simultaneously the
values of \oiii/\Hb, \oii/\Hb and \nii/\Ha within the observational uncertainties. We find solutions
for each value of $z$. In Figure \ref{fig4} we show the acceptable
ranges of \phiholmes/\phitot, $U$, $n_\mathrm{e}$
and N/O as a function of $z$. The models clearly indicate a systematic
decrease of the electron density with increasing $z$. If we restrict to
solar metallicity (dark bars in Fig. \ref{fig4}), we find that the models which
fit the observations become dominated by the HOLMES as $z$
increases. Turning the argument around, this might be an indication
that the metallicity of the DIG is roughly solar. Our models also
indicate that N/O increases with increasing destance to the galactic
plane, at least until $|z| \sim$ 1.5 kpc.

\section{Conclusions}

Our scenario, which considers both the population of OB stars and that
of HOLMES, is able to explain the long standing problem of the
ionization of the DIG (see \cite[Flores-Fajardo et al. (2011)]{2011MNRAS.415.2182F}).

\end{document}